\documentstyle[11pt,newpasp,twoside,epsf]{article}
\reversemarginpar

\begin{document}

\title{Stellar Coronae with \textit{XMM-Newton} RGS\\
       I. Coronal Structure}

\author{M. G\"udel, M. Audard}
\affil{Paul Scherrer Institut, W\"urenlingen and Villigen, CH-5232 Villigen PSI,
       Switzerland}
\author{A.~J. den Boggende, A.~C. Brinkman, J.~W. den Herder, J.~S. Kaastra,
        R. Mewe, A.~J.~J. Raassen, C. de Vries}
\affil{Space Research Organization of the Netherlands, Sorbonnelaan 2, 3584 CA Utrecht,
       The Netherlands}
\author{E. Behar, J. Cottam, S.~M. Kahn, F.~B.~S. Paerels, J.~M. Peterson, 
        A.~P. Rasmussen, M. Sako}
\affil{Columbia Astrophysics Laboratory, Columbia University, 550 West 120th Street, New York, 
        NY 10027, USA}
\author{G. Branduardi-Raymont, I. Sakelliou}
\affil{Mullard Space Science Laboratory, University College London, Holmbury St. Mary,
       Dorking, Surrey, RH5 6NT, United Kingdom}
\author{C. Erd}
\affil{Astrophysics Division, Space Science Department of ESA, ESTEC, 2200 AG Noordwijk, 
       The Netherlands}

\begin{abstract}
First results from  high-resolution coronal spectroscopy with the {\it XMM-Newton} Reflection
Grating Spectrometers (RGS) are reviewed. Five stellar systems (HR~1099, Capella, Procyon,
YY Gem, AB Dor) have been observed. The emphasis of the present paper is on overall coronal
structure. Elemental abundances in {\it active stars} are found to be `anomalous' in the sense that
they tend to increase with increasing First Ionization Potential (FIP - i.e., signifying
an inverse FIP effect). Coronal densities are measured at levels of a few times
$10^{10}$~cm$^{-3}$ for cooler plasma, although there are indications for very high
densities in the hotter plasma components. 
\end{abstract}

\section{Introduction}

\begin{figure}[t!]
\plotone{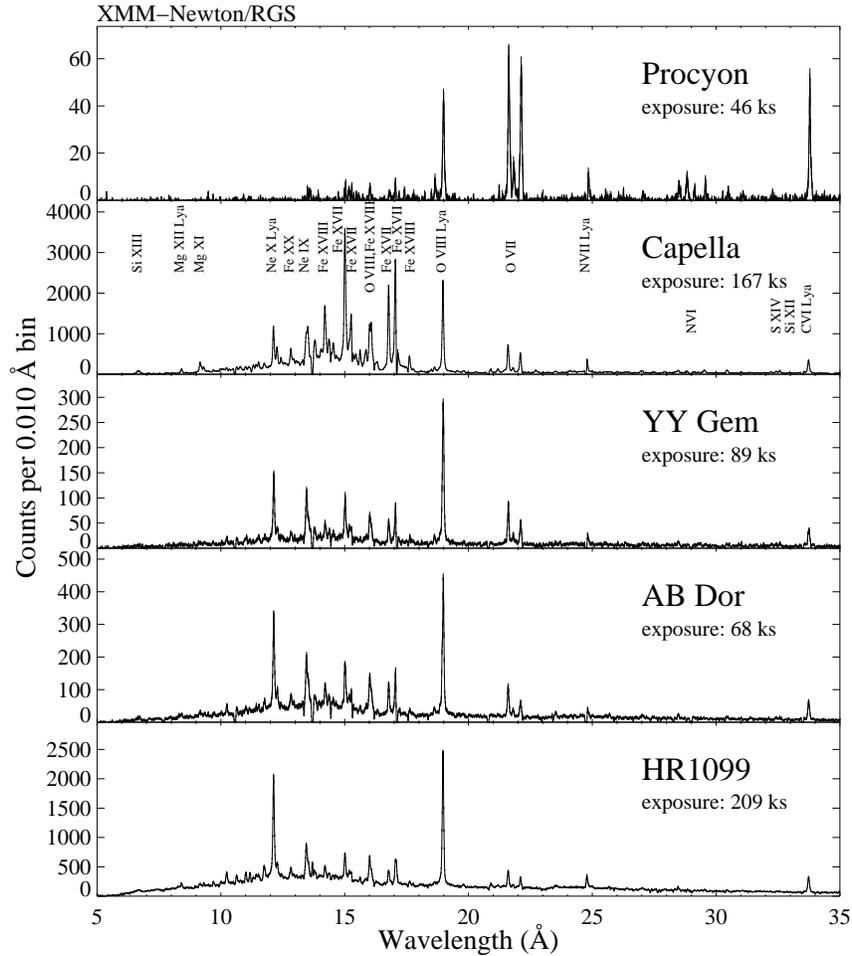}
\caption{Five spectra observed with RGS, smoothed to 30~m\AA\ effective resolution. 
         The average coronal temperature increases from Procyon (top), 
	 to Capella (2nd panel), YY Gem (3rd panel),
	 AB Dor (4th panel) and HR~1099 (bottom).}
\end{figure}

High-resolution X-ray spectroscopy now available from {\it Chandra} and {\it XMM-Newton}
has opened new access to coronal diagnostics. For the first time, densities, elemental
abundances, the thermal structure, and mass motions can be measured explicitly based on
atomic X-ray line diagnostics  for a large number of sources. Such measurements
are pivotal for our understanding of coronal structure, coronal heating, and mass transfer
from the lower layers into the corona. 

The two Reflection Grating Spectrometers (Brinkman et al., this volume) onboard {\it XMM-Newton} 
(Jansen et al., this volume) have started their mission with the First-Light target HR 1099 and
later continued to obtain deep X-ray spectra of various coronal sources for an extended
survey of coronal activity. We report here a first summary of results (see also Audard et al., this volume).

\begin{figure}[t!]
\plotone{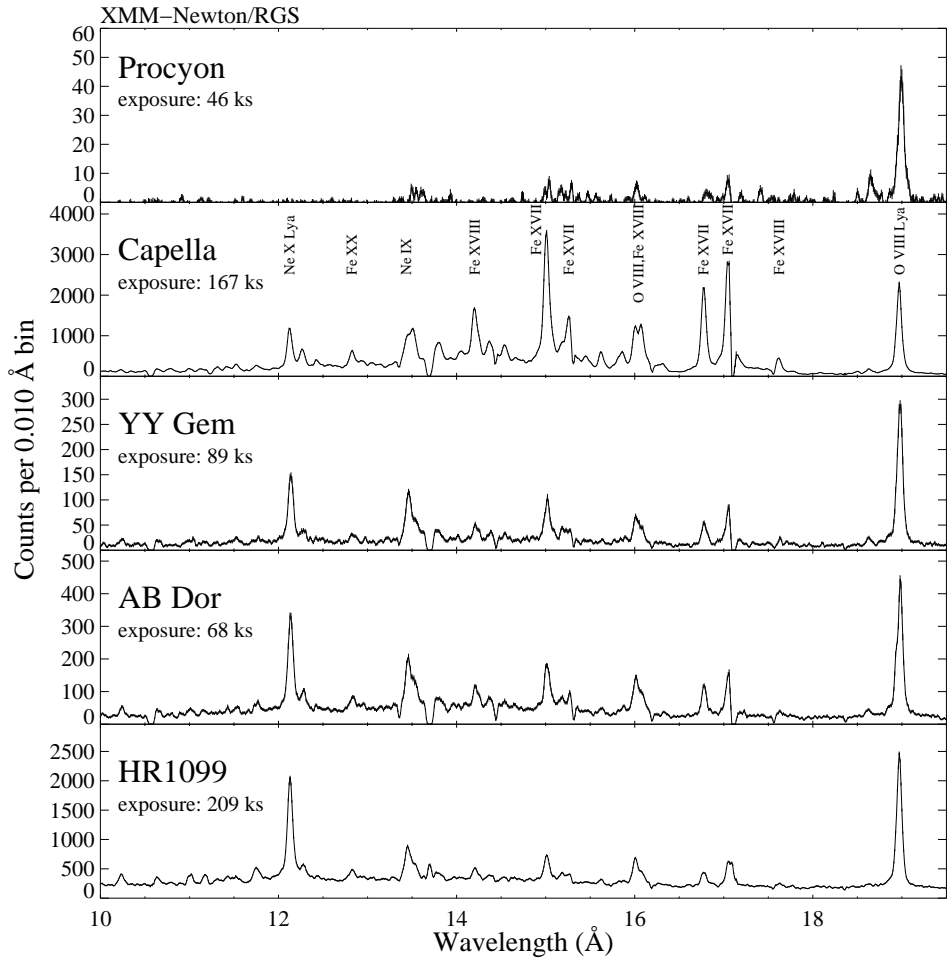}
\caption{Extract from Fig. 1, showing the Fe L-shell region, together with lines of
         Ne and O.}
\end{figure}

\section{RGS Spectra and Thermal Structure}

Stars of higher activity level reveal coronae of higher average temperature. This simple
correlation (see, e.g., G\"udel, Guinan, \& Skinner 1997) has defied a conclusive physical
explanation, although models involving higher average but quiescent heating rate (Schrijver,
Mewe, \& Walter 1984) or higher ``micro''-flare rate (G\"udel et al. 1997) have been discussed in
the literature.

RGS spectroscopy explicitly reveals differences in the thermal structure through atomic lines. 
Figures 1 and 2 show spectra of five sources (Procyon, Capella, YY Gem, AB Dor, and HR 1099 in 
order of increasing `activity' or average temperature). The increasing relative importance of the 
continuum toward higher activity is evident (note that much of Capella's `pseudo-continuum' 
around 15~\AA\ is due to overlapping scattering line wings). The rather cool `solar-like'
corona of Procyon reveals only weak Fe lines, while the spectrum is dominated by
lines  of C~V, N~VI, N~VII, and O~VII, formed around 1--2~MK.
A further evident temperature 
diagnostic is given by the flux ratio between the O~VIII Ly$\alpha$ line at 18.97~\AA\ and 
the O~VII resonance line at 21.6~\AA; clearly, this ratio systematically increases from Procyon 
to HR~1099.

\section{Elemental Abundances}

Coronal elemental abundances are known to deviate from photospheric abundances in the
Sun (Meyer 1985; Feldman et al. 1992) such that elements with a low First Ionization 
Potential (FIP, below 10~eV) are overabundant by an average factor of $\sim$4 (Feldman
et al. 1992). Various scenarios for the related fractionation process have been proposed,
involving magnetic and electric fields (e.g., H\'enoux 1995 for a review). It came
therefore as a surprise when the coronal abundances of {\it active} stars did not show 
evidence for clear FIP dependencies but rather showed that most
elements are {\it depleted} in the corona (Antunes et al. 1994; Drake 1996; White 1996).

Elemental abundances can now unambiguously be determined from separated lines in 
the X-rays. Fig. 3a shows the surprising result (Brinkman et al. 2001) from the RGS 
spectrum of HR~1099 (Fig. 1): the relative abundances (normalized to oxygen) {\it increase} 
with  increasing FIP, in a nearly exponential fashion, with Ne showing an unusually high
abundance evident in the very strong Ne~X line at 12.1~\AA\ in Fig. 1. There is as of
yet no consistent model for this {\it inverse FIP effect}, but Brinkman et al. (2001) 
suggest that mechanisms analogous to solar Ne- and S-rich flares may enrich the corona
with high-FIP species. Similar results are reported for the other two active sources YY Gem
and AB Dor (G\"udel et al. 2001ab). 

\begin{figure}
\plottwo{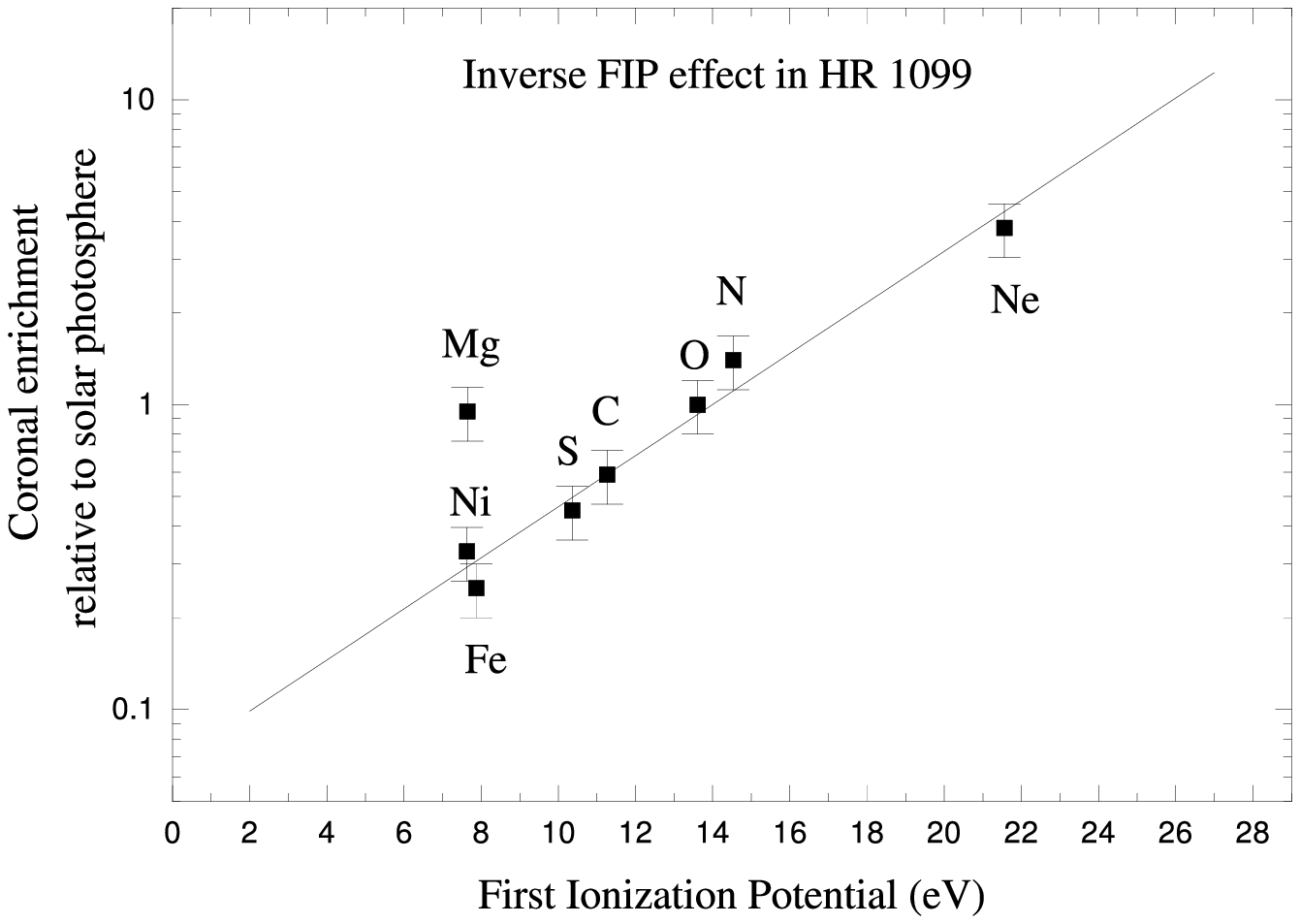}{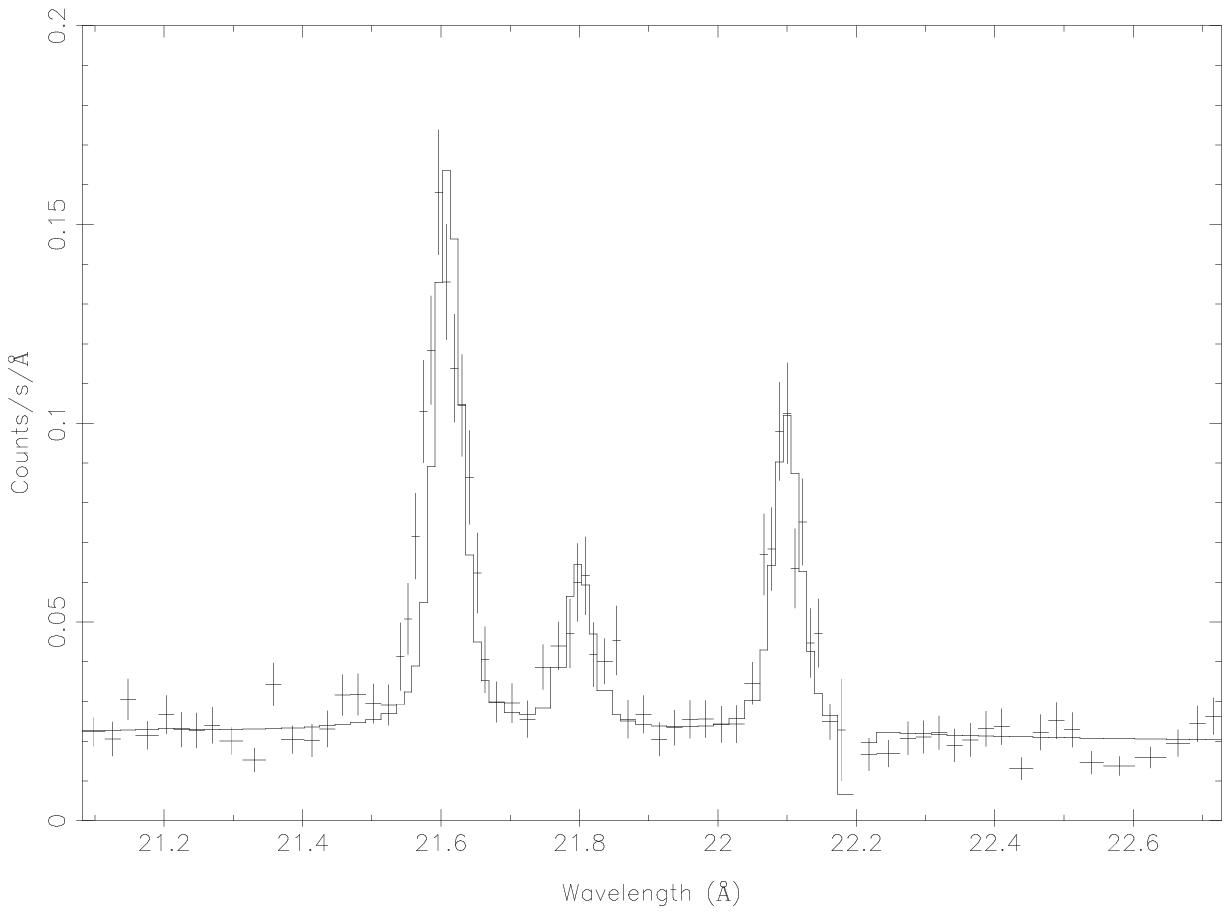}
\caption{{\bf Left:} Coronal abundances derived for HR~1099
  from {\it XMM-Newton} RGS data, relative to solar photospheric, as a function of FIP 
  and normalized to the oxygen abundance. Note the inverse FIP effect (Brinkman et al. 2001). -
 {\bf Right:} Density-sensitive He-like line triplet of O~VII, measured from 
the quiescent AB Dor. The three lines are, for increasing wavelength, the resonance,
the intercombination, and the forbidden line.  The best-fit density is 
$(3\pm 1.5)\times 10^{10}$~cm$^{-3}$ (G\"udel et al. 2001a).}
\end{figure}

\section{Geometric Structure and Densities}

X-ray coronal structure is inferred indirectly, based on eclipse mapping, rotational 
modulation, or density measurements. The latter approach is available in the RGS range 
based on the He-like density-sensitive triplets of N~VI, O~VII, Ne~IX, Mg~XI, and Si~XIII.
Fig. 3b shows an example of the O~VII triplet of the quiescent AB Dor. Best-fit 
measurements indicate a density of $(3\pm 1.5)\times 10^{10}$~cm$^{-3}$, quite
high by solar standards (but possible in solar active regions). This density value refers
to relatively cool material ($T$ of a few MK). However, heavier elements such as Mg tentatively
indicate much higher densities for hotter material, up to the order of $10^{12}$~cm$^{-3}$ 
(Audard et al. 2001 for Capella).  This bi-modal distribution of densities 
(cool material at low densities, hot material at higher densities) suggests the
presence of  separate
loop structures, with low-lying, compact hot loops on the one hand and larger, cooler loops
on the other hand (Mewe et al. 2001).

The luxury of binary stellar eclipses is rarely available, but when it is, the 
mutual eclipses can be used to infer the geometry of the X-ray corona. The dM1e active
binary system YY Gem was extensively observed by {\it XMM-Newton} (G\"udel et al. 
2001b), including three deep eclipses. The eclipse light curves were used to reconstruct
realisations of fitting geometric coronal models. It was found that the plasma is required
to be relatively close to the star, with an average density scale height of
$(1-4)\times 10^{10}$~cm (Fig. 4a), which is fully compatible with the spectroscopically 
measured $T$ of  3$-$10~MK. Most of the detected X-ray light is then emitted
from densities around several times $10^9$~cm$^{-3}$ to a few times $10^{10}$~cm$^{-3}$
(Fig. 4b), and these values are again in gratifying agreement with spectroscopic results.

\begin{figure}
\plottwo{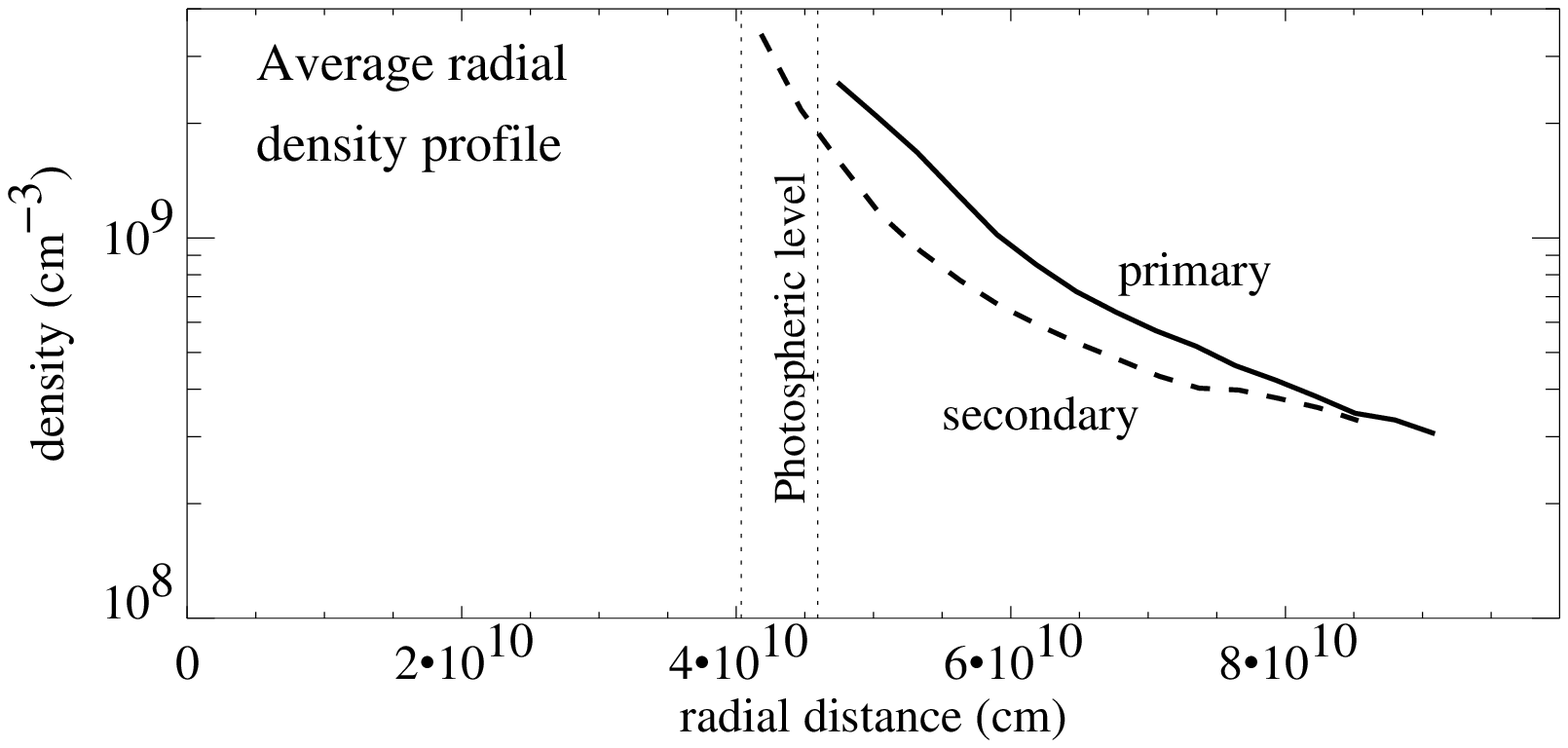}{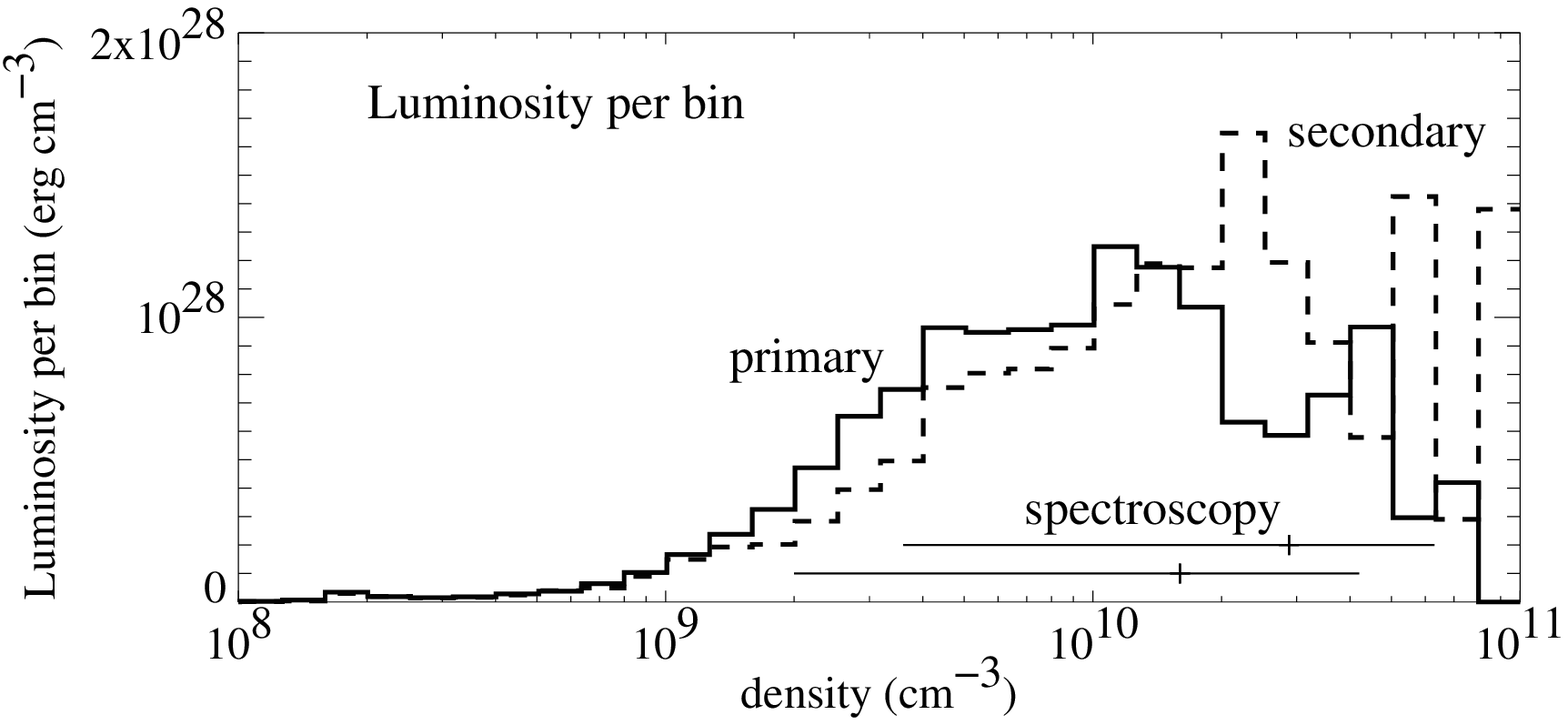}
\caption{Coronal modeling from light curve inversion of the eclipsing binary YY Gem.
{\bf Left:} Average radial density profile for each component. {\bf Right:}
 Luminosity contribution as a function of density. The bars at the bottom indicate
 the spectroscopically measured electron densities from RGS data  (two methods,
  see G\"udel et al. 2001b). }
\end{figure}

\section{Summary and Conclusions}

Stellar coronae have been compared with scaled-up versions of the solar corona
although this analogy has clearly become doubtful. Not only are the X-ray luminosities
of active stars orders of magnitude larger than the Sun's, but the thermal
structure is  also very different, ranging from cooler solar-like structures
of a few MK (probably analogous to the Procyon corona)
up to {\it non-flaring} coronal components at 20$-$30~MK. 
Spectroscopy clearly provides the necessary information for further study.

The most crucial results from this early {\it XMM-Newton} RGS stellar coronal survey
set active stars apart from the Sun. Coronal elemental abundances, although known before
to deviate from solar behavior, are now showing a clear systematic dependence on
the FIP value, such that high-FIP elements (O, N, Ne) are largely overabundant compared to
low-FIP elements (Fe, Mg, Si, Ni), thus defining an {\it inverse FIP effect}, contrary to 
the solar trends. The Sun does occasionally reveal Ne- and S-rich flares (i.e., high-FIP 
elements are enriched in the flaring plasma), and analogous processes may be 
responsible for the anomalously high Ne abundance seen in active stars. 

Densities found so far for active stars are high for solar standards, even for relatively
cool plasma. High densities in hot plasma, although still only marginally detected,  
have no solar analogy except for flares. Taken together, we speculate that the mechanisms 
for heating to high temperatures and for inverting the coronal abundance pattern 
are related, but unknown on the {\it non-flaring} Sun.

\acknowledgments
We thank all {\it XMM-Newton} teams for their support in the evaluation of the present and
related data. The PSI group is supported
by the Swiss Academy of Natural Sciences and the Swiss National Science
Foundation (grants 2100-049343 and  2000-058827). SRON is financially
supported by the Netherlands Organization for Scientific Research (NWO).
The Columbia University team acknowledges generous support from the
National Aeronautics and Space Administration.  MSSL acknowledges support 
from the Particle Physics and
Astronomy Research Council. This work is based on observations obtained with
XMM-Newton, an ESA science mission with instruments and contributions directly
funded by ESA member states and the USA (NASA).

\end{document}